\documentclass[%
 reprint,
 amsmath,amssymb,
 aps,
prb,
]{revtex4-2}

\usepackage{graphicx}
\usepackage{dcolumn}
\usepackage{bm}
\usepackage[version=4]{mhchem}

\begin{document}

\title{Switching of ferrotoroidal domains via an intermediate mixed state in the multiferroic Y-type hexaferrite Ba$_{0.5}$Sr$_{1.5}$Mg$_{2}$Fe$_{12}$O$_{22}$}

\author{Jiahao Chen}
\affiliation{Department of Physics, University of Oxford, Oxford OX1 3PU, England, United Kingdom}

\author{Francis Chmiel}
\affiliation{Department of Physics, University of Oxford, Oxford OX1 3PU, England, United Kingdom}

\author{Roger D. Johnson}
\affiliation{Department of Physics and Astronomy, University College London, Gower Street, London WC1E 6BT, United Kingdom}
\affiliation{Department of Physics, University of Oxford, Oxford OX1 3PU, England, United Kingdom}

\author{Jieyi Liu}
\affiliation{Department of Physics, University of Oxford, Oxford OX1 3PU, England, United Kingdom}

\author{Dharmalingam Prabhakaran}
\affiliation{Department of Physics, University of Oxford, Oxford OX1 3PU, England, United Kingdom}

\author{Paolo G. Radaelli}
\email{paolo.radaelli@physics.ox.ac.uk}
\affiliation{Department of Physics, University of Oxford, Oxford OX1 3PU, England, United Kingdom}

\begin{abstract}

We report a detailed study of the magnetic field switching of ferrotoroidal/multiferroic domains in the Y-type hexaferrite compound \ce{Ba_{0.5}Sr_{1.5}Mg_{2}Fe_{12}O_{22}}. By combining data from SQUID magnetometry, magneto-current measurements, and resonant X-ray scattering experiments, we arrive at a complete description of the deterministic switching, which involves the formation of a temperature-dependent mixed state in low magnetic fields. This mechanism is likely to be shared by other members of the hexaferrite family, and presents a challenge for the development of high-speed read-write memory devices based on these materials.

\end{abstract}

\maketitle


\section{Introduction}

Harnessing the intrinsic functionality of magneto-electric (ME) and multiferroic (MF) materials in which (anti)ferromagnetic order can be controlled by electric fields, and conversely ferroelectric order controlled by magnetic fields \cite{eerenstein2006multiferroic,kimura2012magnetoelectric,nan2008multiferroic,hill2000there}, may enable the development of high-speed, low energy magnetic storage devices. A particularly promising group of materials is the hexaferrite family, which includes several ``type-II" multiferroics with significant electric polarisation induced via the inverse-Dzyaloshinskii-Moria interaction (i-DMI) upon non-collinear magnetic ordering \cite{katsura2005spin,mostovoy2006ferroelectricity}. Remarkably low fields are required for switching, and this class of materials could well be integrated into in MF storage devices provided ordering temperatures are raised above room temperature \cite{kocsis2019magnetization,kitagawa2010low}.

Within the wider hexaferrite family, Y-type hexaferrites were the first found to show significant ME effects, and have remained of great interest given that their low-temperature ME coefficients are among the largest observed in any material. The Y-type compound \ce{Ba_{0.5}Sr_{1.5}Mg_{2}Fe_{12}O_{22}} (BSMFO) discussed in this paper displays switching behaviour typical of most Y-type hexaferrites: at low temperatures and under moderate magnetic fields, BSMFO adopts a complex magnetic structure, known as `two-fan transverse conical' (TFTC - see below), which has a sizeable ferrimagnetic moment \textbf{M} and a magnetically-induced polarisation \textbf{P} that is orthogonal to \textbf{M} and to the 6-fold axis of the crystal structure. The development and switching of the macroscopic polarisation under remarkably low magnetic fields are most readily described in terms of a ferrotoroidal order parameter $\mathbf{T}\propto\mathbf{P}\times\mathbf{M}$, which has become known as the fourth primary ferroic order parameter \cite{van2007observation,sawada2005lorentz,fiebig2009current}. In essence, cooling in a cross-field configuration tends to stabilise a single or at least a dominant TFTC toroidal domain (which we will refer to as the `mono-domain state'), as demonstrated in our previous work (Figure 7 in reference \citenum{Chmiel2019Magnetoelectric}). The ferrotoroidal domain population is unaffected by subsequent reversals of the magnetic field at low temperatures, except very near $B=0$ (see below). Indeed, reversing the magnetic or electric fields alone will not change the sign of the ferrotoroidicity, which can only be reversed by the application of cross fields. This behaviour leads to the deterministic switching of the electric polarisation upon reversal of the sample's magnetisation for a given, preconditioned, ferrotoroidal state. At the microscopic level, within a single ferrotoroidal domain, it was proposed that reversing the magnetic field led to a coherent rotation of the TFTC structure about an in-plane axis, such that the deterministic switching is mediated via an intermediate longitudinal-conical (LC) phase, defined below \cite{ishiwata2008low}. We later showed that the rotation actually occurs about the $c$ axis, and therefore not via an LC phase \cite{Chmiel2019Magnetoelectric}. Despite this progress, anomalous temperature-dependent features in the magnetisation, electric polarisation, and resonant X-ray scattering (RXMS) intensities observed upon domain switching at low fields had not been explained \cite{2008Ferroelectric, Hajime2009Two,Chmiel2019Magnetoelectric}, but were thought to indicate a more complex switching mechanism; perhaps a two-step process rather than a simple rotation.

In this paper, we present a combined analysis of magnetometry, ME measurements, and linearly polarised RXMS to show that, while an intermediate longitudinal-conical phase does not mediate ferrotoroidal domain switching, its presence in a mixed state at low fields can account for all anomalous features observed. Hence, we are now able to provide a fully comprehensive description of deterministic multiferroic domain switching in BSMFO based on a simple rotation of the TFTC structure within ferrotoridal domains of a unique and controllable sign.

\section{Crystal and magnetic structures}

\begin{figure}
  \centering
  \includegraphics[width=8.7cm]{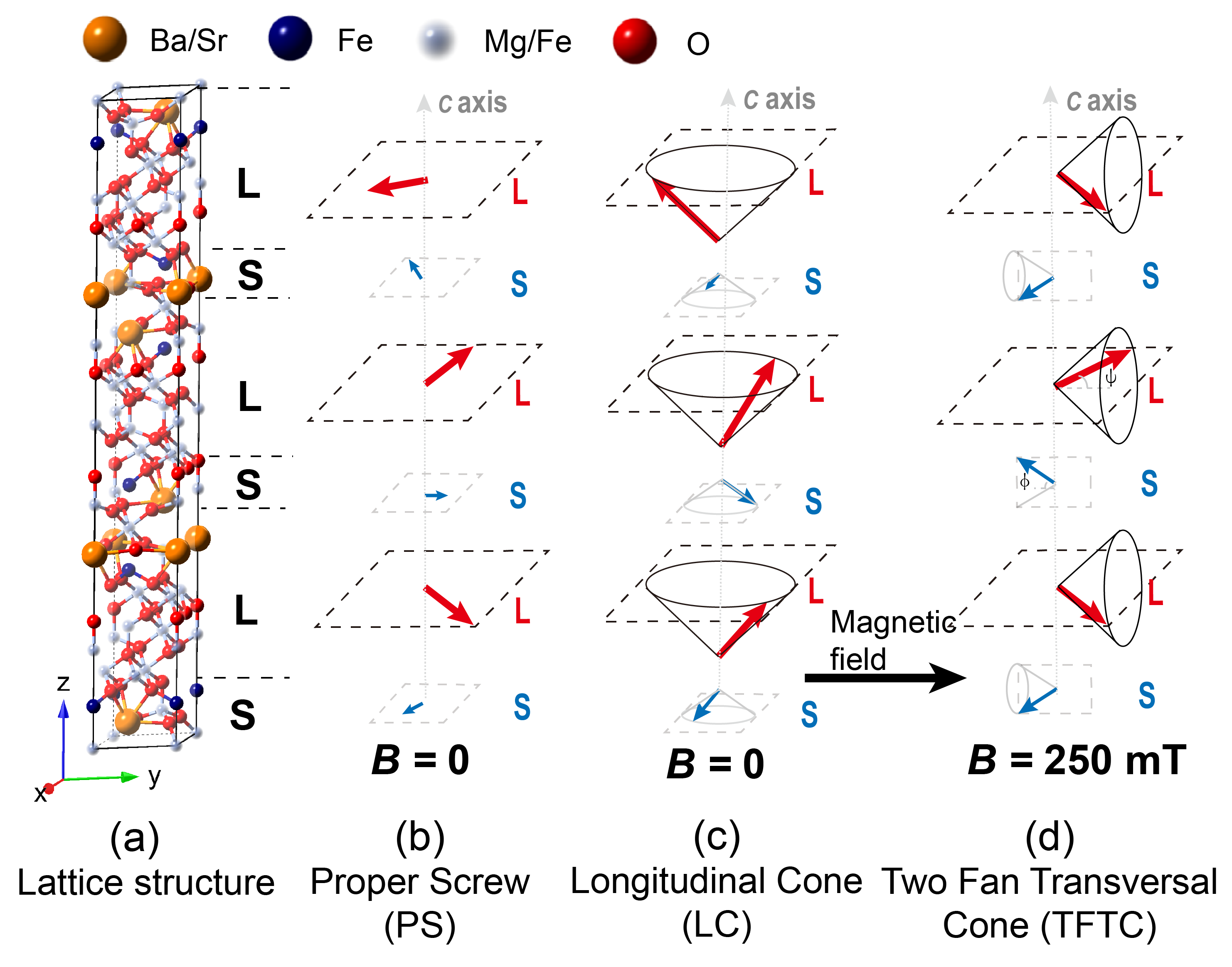}
  \caption{The crystal structure (a) and magnetic structures (b)-(d) of Y-type hexaferrites.  The magnetic structures are internally complex but are depicted using the so-called `block model' approximation (see text), in which the hexagonal unit cell is divided into quasi-collinear ferrimagnetic blocks stacked along the $c$ axis. (b) At room temperature the magnetic structure is described by a planar helix in the $ab$ plane. (c) On cooling, the spins cant slightly towards the $c$ axis and the longitudinal conical (LC) structure emerges with a small net ferrimagnetic moment parallel to the $c$ axis (less than 1 $\mu_B$/f.u.)\cite{Hajime2009Two,ishiwata2010neutron,Koksis_2020}. (d) Upon application of a magnetic field in the $ab$ plane, the cone `flops' so that its axis, and ferrimagnetic magnetisation, is directed along the field, leading to the two-fan transverse cone (TFTC) magnetic structure.}
  \label{fig:1_structure}
\end{figure}

\ce{Ba_{0.5}Sr_{1.5}Mg_{2}Fe_{12}O_{22}} (BSMFO) crystallises in the $R\bar{3}m$ space group, with lattice parameters $a = b \approx 5.8$  \AA \ and $c \approx 43$ \AA. The primitive unit cell contains six symmetry-inequivalent Fe ions, four in octahedral coordination and two in tetrahedral coordination. A simple model of the magnetic structure can be obtained by grouping the Fe sites into quasi-collinear ferrimagnetic blocks, whereby the intra-block exchange interactions are largely unfrustrated and predominantly antiferromagnetic. By contrast, the inter-block interactions are severely frustrated by geometry, which is the main cause of the hexaferrites' complex magnetic phase diagram \cite{utsumi2007superexchange}. This simplification leads to the so called ``block model", in which all the spins in each block are added together into a ``super-spin" proportional to the block's ferrimagnetic moment. This approximation can be used to construct phenomenological models, and provide a reasonable approximation for the neutron scattering structure factor used to determine the magnetic structure of each phase \cite{momozawa1993field,lee2012heliconical,Momozawa2001Cation}. The simplest version of the block model is the two-block model, in which block L (large ``super-spin") and block S (small ``super-spin") are taken as the basic units of the magnetic structure.

Figure \ref{fig:1_structure} shows the crystal structure (Figure \ref{fig:1_structure}(a)) and three two-block magnetic structures (Figure \ref{fig:1_structure}(b)(c)(d)), which are most relevant for Y-type hexaferrite \ce{BSMFO}. At high temperatures, \ce{BSMFO} is ferrimagnetic, while at room temperature and zero magnetic field it exhibits a planar helical structure with proper-screw axis along the \textit{c} axis and magnetic propagation vector $k_0$ = (0,0,0.9), as shown in Figure \ref{fig:1_structure}(b). On cooling below 50 K, a longitudinal cone (LC) structure with an out-of-plane component of the moment is formed \cite{ishiwata2010neutron,nakamura2012mossbauer}, as shown in Figure 1(c), giving rise to a bulk ferrimagnetic moment parallel to the $c$ axis. Upon the application of a small magnetic field within the $ab$ plane, the ferrimagnetic magnetisation, and hence axes of the cone, rotates to coincide with the field direction, forming the so called transverse cone (TC) magnetic structure \cite{zhai2017giant,ishiwata2008low}. Although different variants of the TC structures are reported throughout the Y-type hexaferrite family, the one relevant to \ce{BSMFO} is the two-fan TC phase (TFTC), described by two magnetic propagation vectors $k_1$ = (0,0,0) and $k_2$ = (0,0,1.5) \cite{kimura2012magnetoelectric}, whereby $k_1$ describes the $\Gamma$ point component of the two block structure giving rise to the ferrimagnetic magnetisation, while $k_2$ describes the oscillating `fan' component. The TFTC phase is ferroelectric due to the i-DMI effect, the polarisation being within the $ab$ plane and perpendicular to the ferrimagnetic magnetisation (i.e. to the applied magnetic field). By contrast, the in-plane incommensurate helical phase and the LC phase are non-polar and therefore have zero polarisation.

\section{Experimental techniques}

The in-plane electric polarisation was measured as a function of perpendicular in-plane magnetic field at fixed temperature using the magneto-current technique. A single crystal sample was prepared in a parallel-plate capacitor geometry, and mounted on a custom probe within a Quantum Design PPMS. Magnetic fields were swept at a fixed rate of 5 mT/s while logging the perpendicular magneto-current, which was integrated to obtain the electric polarisation, \textbf{P}. The in-plane magnetisation was measured using a Quantum Design MPMS-3 magnetometer, with the magnetic field oriented along the $a^*$ axis of the single crystal (the [2,1,0] direction in real space). Field sweeps at fixed temperature (MvH) were performed at 10, 25, 50, and 125 K. We found that, in general, the transitions observed in the magnetisation were significantly \emph{sharper} than those observed in the magneto-current, likely due to the fact that the polarisation switching curve is broadened both by the sweep rate of the magnetic field and by the impedance of the capacitive/inductive component in the effective circuit. We also note that the accuracy of the above magneto-current measurements depends both on the conductivity of the sample (which is very low at low temperature but increases on warming) and on the background noise. 

RXMS measurements at the Fe $L_{3}$ edge (707 eV) were performed on the RASOR diffractometer at beamline I10, Diamond Light Source (UK). A cleaved single crystal sample was mounted with the $(00l)$ zone axis surface normal, on the end of a $^4$He cryostat equipped with a small resistive electromagnet capable of applying $\pm 250$ mT within the plane of the sample (full details are given in reference \citenum{Chmiel2019Magnetoelectric}). $\theta/2\theta$ scans that spanned the $(0,0,3)$ Bragg peak and its magnetic satellites were performed at different magnetic fields and temperatures, with the helical and TFTC phases identified by their characteristic propagation vectors (satellite positions) of $(0,0,\pm0.9)$ and $(0,0,\pm1.5)$, respectively. The presence of circular dichroism (normalized difference in diffraction intensity for right- and left-handed circularly polarised light) provided quick confirmation of the magnetic origin of the satellite intensities. No circular dichroism was observed at $(0,0,3)$ in the helical phase, while for the TFTC phase this peak displayed weak circular dichroism since it includes a significant contribution from the $\Gamma$ point ferrimagnetic order. Magnetic moment vectors of a given Fourier component of each magnetic structure could be decomposed using linear polarisation analysis of the diffraction intensity, which was measured in four channels; $\sigma-\sigma'$,$\sigma-\pi'$, $\pi-\sigma'$, and $\pi-\pi'$, where $\sigma$ denotes the polarisation vector perpendicular to the scattering plane, $\pi$ denotes the polarisation vector within the scattering plane, and the prime indicates the scattered radiation. Detailed calculations of the linear polarisation dependence on the diffraction intensities can be found in reference \cite{Chmiel2019Magnetoelectric}. 

\begin{figure}
  \centering
  \includegraphics[width=8.8cm]{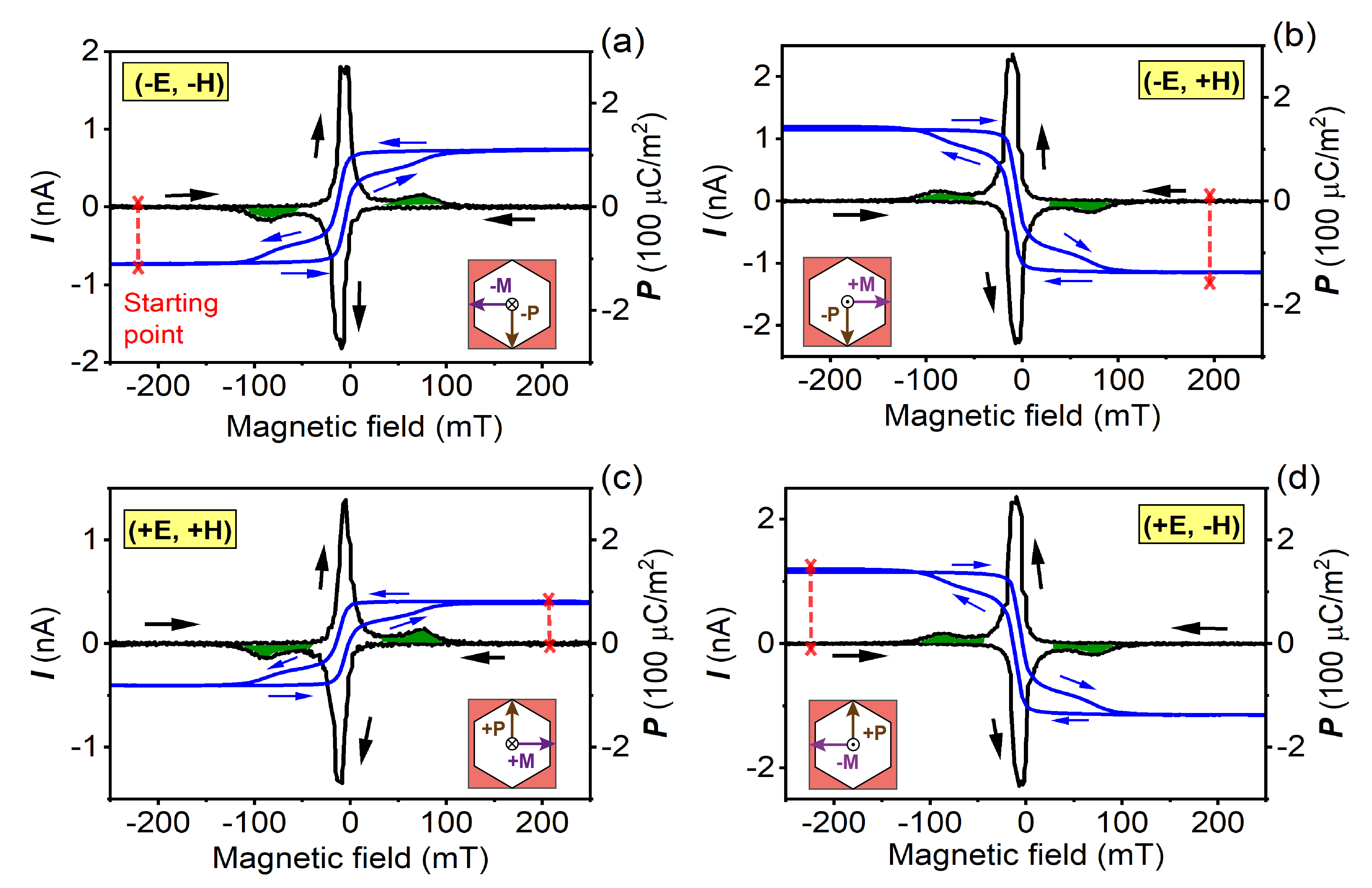}
  \caption{Two-stage magnetic-field-driven electrical polarisation switching of BSFMO at 10 K. (a)-(d) Current-field (\textit{I-H}) and polarisation-field (\textit{P-H}) hysteresis loops, based on magneto-current measurement collected after cooling with poling electric and magnetic fields of different orientations (see boxes): \textbf{(a)} (-E,-H) field; \textbf{(b)} (-E,+H) field; \textbf{(c)} (+E,+H) field; \textbf{(d)} (+E,-H) field. The poling fields were removed after cooling and prior to measuring.  The starting point of each magnetic field sweep is indicated by red crosses. The second, smaller peak associated with the second stage of switching is emphasised by a dark (green) filling. The insets show the directions of the initial polarisation \textbf{P} and magnetisation \textbf{M} after poling, and the resulting direction of the toroidal moment along the $c$ axis, which is never changed during the magnetoelectric switching process.}
  \label{fig:2_toroidal}
\end{figure}

\section{Results and discussion}

\subsection{Ferrotoroidal domains}

The measurements reported in the following sections were performed having preconditioned the sample into a monodomain state by cross-field cooling. The field poling process began at 200 K when the sample is sufficiently insulating to maintain the bias electric field. A magnetic field \textit{H} of $\pm$250 mT was applied along the $a$ axis (deep into the TFTC phase at low temperature), with a $\pm$100 V bias voltage simultaneously applied along $b^*$ (i.e. orthogonal to $a$ within $ab$ plane). The sample was cooled to 10 K and then the electric field bias was removed. Depending on the sign of the two fields one of four combinations of \textbf{P} and \textbf{M} could be established; $\{-\mathbf{P}|-\mathbf{M}\}$, $\{-\mathbf{P}|+\mathbf{M}\}$, $\{+\mathbf{P}|+\mathbf{M}\}$, or $\{+\mathbf{P}|-\mathbf{M}\}$ (see Figure \ref{fig:2_toroidal} a to d).

The conservation of ferrotoroidicity through magnetic field reversals is clearly demonstrated in Figure \ref{fig:2_toroidal}. Upon reversing the magnetic field at low temperatures, a current in the orthogonal direction (black line) is induced due to the reversal of polarisation. Furthermore, the full magnitude of the polarisation (blue lines) is completely recovered at high fields after reversal, regardless of the initial configuration, and is retained even after multiple cycles. These observations are consistent with the ferrotoroidal domain population always being restored at high applied magnetic fields, while both \textbf{P} and \textbf{M} simultaneously switch upon magnetic field reversal. 

The magneto-current data and resultant \textit{P-H} loops indicate that there is anomalous behaviour in the low-field region where switching occurs. For example, starting from the (-E, -H) state (in panel \textbf{a}), a sharp positive peak emerges when the field approaches zero, followed by a second, smaller positive peak at around +75 mT, giving rise to a low-field plateau in the electric polarisation within an apparent 2-step switching process. By reversing the direction of the sweep, the pattern is very similar but all signs are reversed, with a first large peak at B = 0 and smaller peak near -75 mT. Different bias field configurations in Figure \ref{fig:2_toroidal} (b-d) yield the same qualitative picture, though the magnitude of the saturated polarisation varies slightly, implying a small bias towards one ferrotoroidal domain over the other. The reason for this bias is still unknown but similar effects in electric field poling were identified in \ce{Ba_{0.4}Sr_{1.6}Mg_{2}Fe_{12}O_{22}} and attributed to clamping between ferroelectric/two-fan cone domains \cite{zhai2018electric, zhai2017giant}.

\subsection{Low-field intermediate mixed state}

\begin{figure}
  \includegraphics[width=8.8cm]{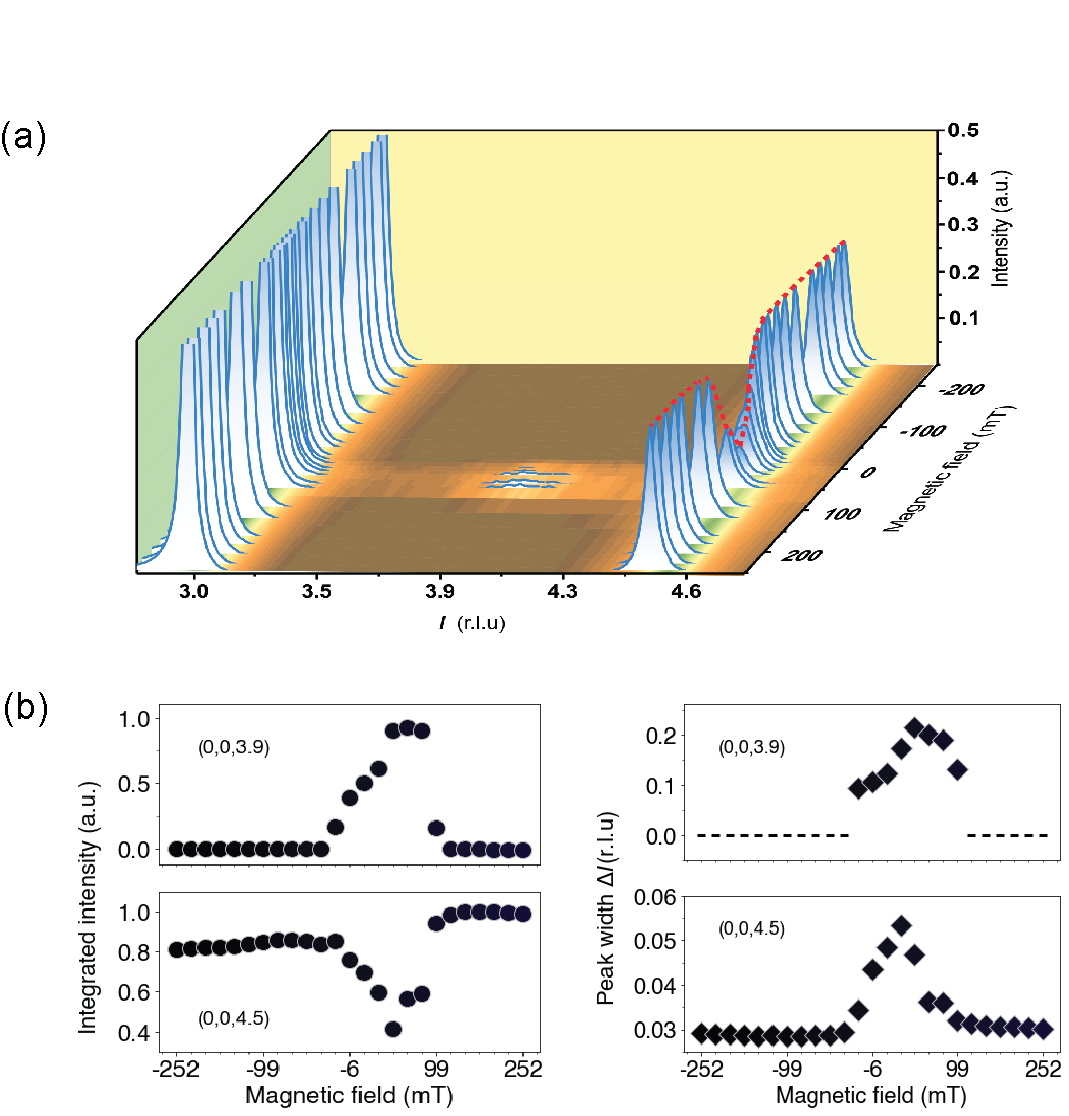}
  \caption{TFTC/LC phase coexistence through the magnetoelectric switching of BSFMO, as determined from RXMS data collected at low temperature along the $(0,0,l)$ zone axis. (a). Peaks at (0,0,3) (\textbf{left}), (0,0,3.9) (\textbf{center}) and (0,0,4.5) (\textbf{right}) correspond, respectively, to the $\Gamma$ point (both phases, Thomson and magnetic scattering), the incommensurate magnetic peak of the LC phase and the commensurate $Z$-point magnetic peak of the TFTC phase. The red line is a guide to the eye. (b). Intensity and FWHM (full width at half maximum) as a function of magnetic field of the commensurate (0,0,4.5) and incommensurate (0,0,3.9) peaks, extracted from Lorentzian fits on peaks by subtracting the background. The dashed line indicates the field region where the intensity of the (0,0,3.9) ICM peak is zero.}
  \label{fig:3_field_scan}
\end{figure}

RXMS was used to investigate the microscopic origin of the anomalous behaviour observed at low fields during switching at low temperature. We recall that the RXMS signatures of the TFTC phase are the purely magnetic commensurate satellites at (0,0,1.5) and (0,0,4.5) (here only the latter was measured), while the (0,0,3) Bragg peak acquires a magnetic contribution. Figure \ref{fig:3_field_scan} shows the evolution of the diffraction pattern as a function of magnetic field. As the magnetic field is swept from -250 mT towards zero, we observe a decrease in the intensity of the (0,0,4.5) satellite starting from around -10 mT, while a broad incommensurate peak at (0,0,3.9) emerges, roughly corresponding to the propagation vector of the LC phase. We note that the large width of the (0,0,3.9) peak indicates that this incommensurate magnetic order has a relatively short correlation length. As highlighted by the dotted guide line, the (0,0,4.5) peak intensity remains low in the field range between -6 mT and 100 mT, throughout which the ICM peak persists. \textbf{Interestingly, in this field regime, the width of the (0,0,4.5) peak also increases, indicating a shorter correlation length, most likely associated to the formation of ICM domains that break up larger CM domains.} Above 100 mT the ICM peak disappears and the (0,0,4.5) peak intensity quickly recovers. The (0,0,3) peak (not shown in full) also exhibits a behaviour indicative of phase coexistence in the low-field range, which will be discussed in more detail in section \ref{sec:RXMS_linear}. Similar measurements performed on sweeping the magnetic field from +250 mT to -250 mT (not shown here) showed the same qualitative behaviour, with the ICM peak appearing in the field range +6 mT and -100 mT. If we assume that the diffraction intensities characteristic of the TFTC and LC structures are proportional to the respective phase fractions, our RXMS results are consistent with an intermediate mixed TFTC/LC state occurring in magnetic fields that exactly coincide with the anomalous features observed in the electric polarisation switching. Considering these data together, we can propose the following scenario: as the field passes through zero, the TFTC phase rotates rapidly, but is also partially converted to the non-polar ICM LC phase (N.B. this is the only phase observed upon cooling in zero magnetic field). The mixed state persists up to 100 mT, when the non-polar LC phase disappears and a pure TFTC state is fully recovered in the same ferrotoroidal domain as at -250 mT. This results in a sudden increase of the polarisation at $\sim$ 100 mT, consistent with the second 'peak' observed in the magneto-current measurements.

\subsection{Field-temperature phase diagram}

\begin{figure}[]
  \includegraphics[width=8.7cm]{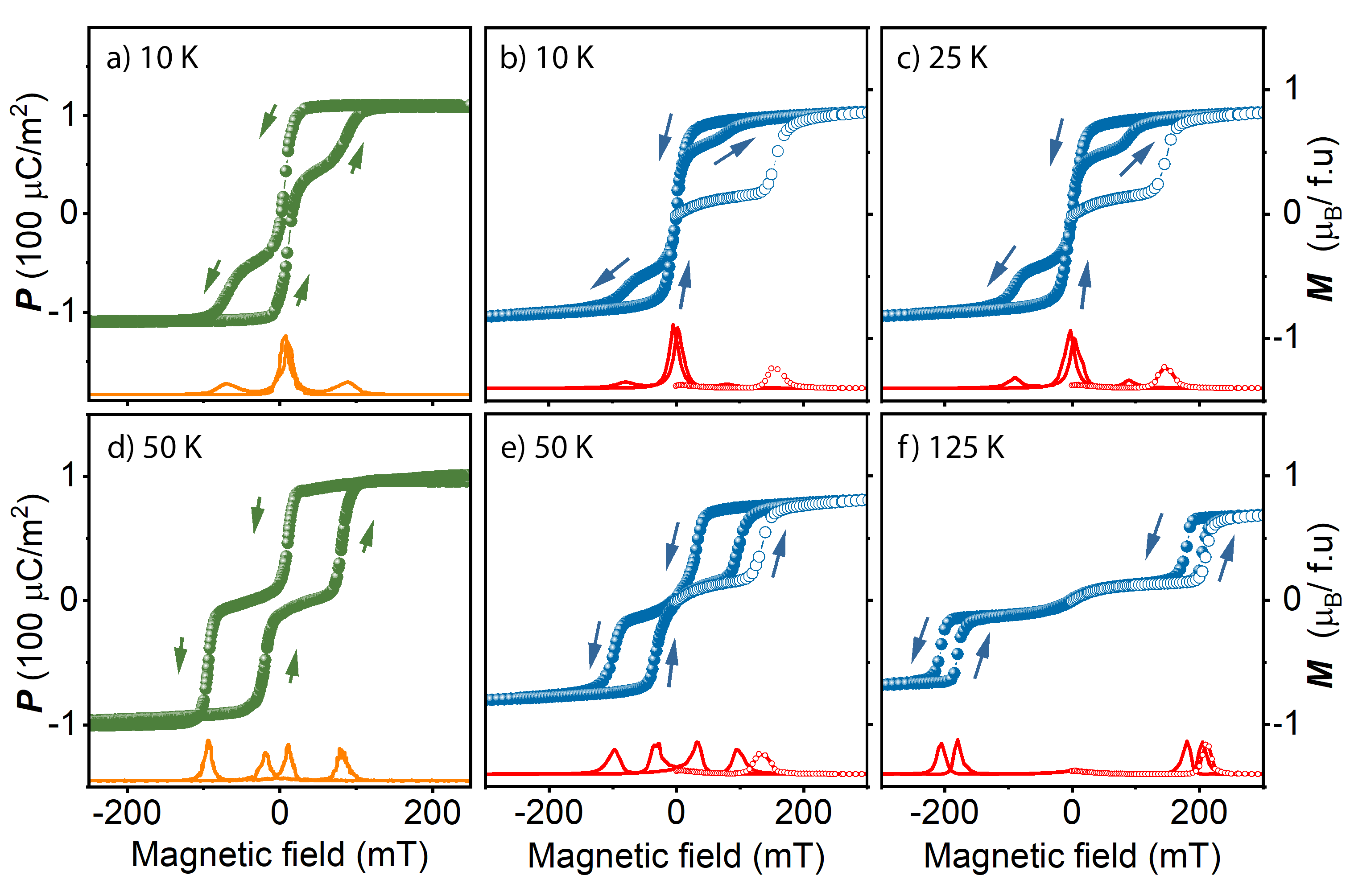}
  \caption{(Colour online) Magnetoelectric switching of BSFMO hexaferrite at different temperatures.  (b,c,e, and f) MvH loops and (a and d) PvH loops, measured at different temperatures with the magnetic field applied along the along the $a$ axis and the polarisation measured along the $b^*$ direction. Open/closed symbols correspond to virgin/subsequent cycles, respectively (see arrows for the field directions).  A field-derivative curve is provided at the bottom for convenience.  In panels (a) and (d), the bottom curve (in yellow) actually corresponds to the absolute value of the magneto-current data, which are integrated to yield the polarisation (green).
  }
  \label{fig:4_SQUID}
\end{figure}

The field dependent magnetisation (MvH) was measured at 10 K and 50 K, and is shown in Figure \ref{fig:4_SQUID} (panels b and e) alongside the respective PvH response (panels a and d). Intermediate plateaus between two steps in the PvH curves, which were found to originate in the intermediate mixed state, are mirrored in the MvH data. This observation is consistent with the expectation that only the TFTC phase has a finite magnetisation within the hexagonal basal plane, which is reduced (in the bulk) throughout the mixed state. In this case, the magnetisation may serve as an indirect measure of the TFTC domain population. The remaining panels in Figure \ref{fig:4_SQUID} (c and f) show the MvH data for an intermediate temperature (25 K) and a higher temperature (125 K). At all temperatures a zero-field step is present regardless of the sweeping directions, followed by a second hysteretic step at higher fields. However, the relative height of the zero-field step decreases upon warming, while the relative height of the hysteretic second step increases and is pushed to higher fields. The hysteresis of the second steps also narrows at high temperatures, indicating a reduced coercivity. These observations imply that on warming, deterministic switching of the ferroelectric polarisation due to the rotation of the TFTC phase is maintained at zero-field, but in a reduced ferrotoridal domain fraction, while the phase fraction of the intermediate mixed state increases correspondingly.

\begin{figure}[]
  \includegraphics[width=0.5 \textwidth]{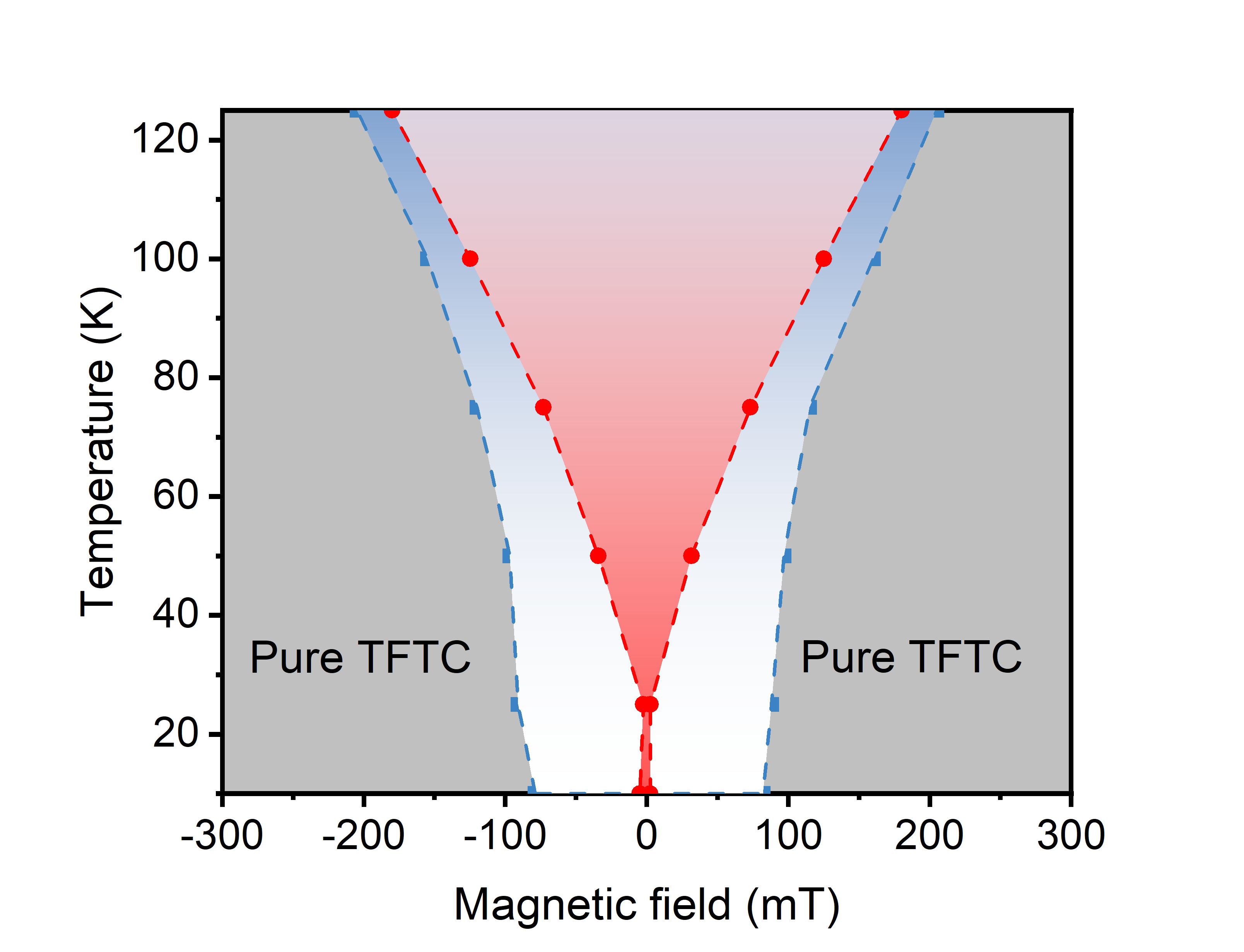}
  \caption{Field-temperature phase diagram of BSFMO. Red and blue symbols connected by dashed lines are extracted from magnetisation data (see Fig. \ref{fig:4_SQUID}) and indicate the fields at which the secondary hysteretic steps occur (see text). The central red shaded area correspond to the region where TFTC/LC phase coexistence always occurs, while the light blue shaded area is the hysteretic region. In the solid grey area only the TFTC phase is observed.
      }
  \label{fig:5_phase diagram}
\end{figure}

The field-temperature phase diagram shown in Figure \ref{fig:5_phase diagram} was derived from a series of MvH measurements at different temperatures up to 120 K (the maximum temperature was limited to ensure accurate electric field poling). The red and blue dots connected by dashed lines of the same colors correspond, respectively, to the positions of the low-field and high-field steps (peaks in the first derivative of the magnetisation -- see Fig. \ref{fig:4_SQUID}), with the area in between (shaded in light blue) representing the hysteretic region. The TFTC structure is always fully restored at high field, as shown in gray filling, and the region where the LC and TFTC structures always coexist in a mixed state regardless of history is shown in red.

\subsection{Modelling RXMS data through the field switching transition}\label{sec:RXMS_linear}

\begin{figure}
	\includegraphics[width= 8.4cm]{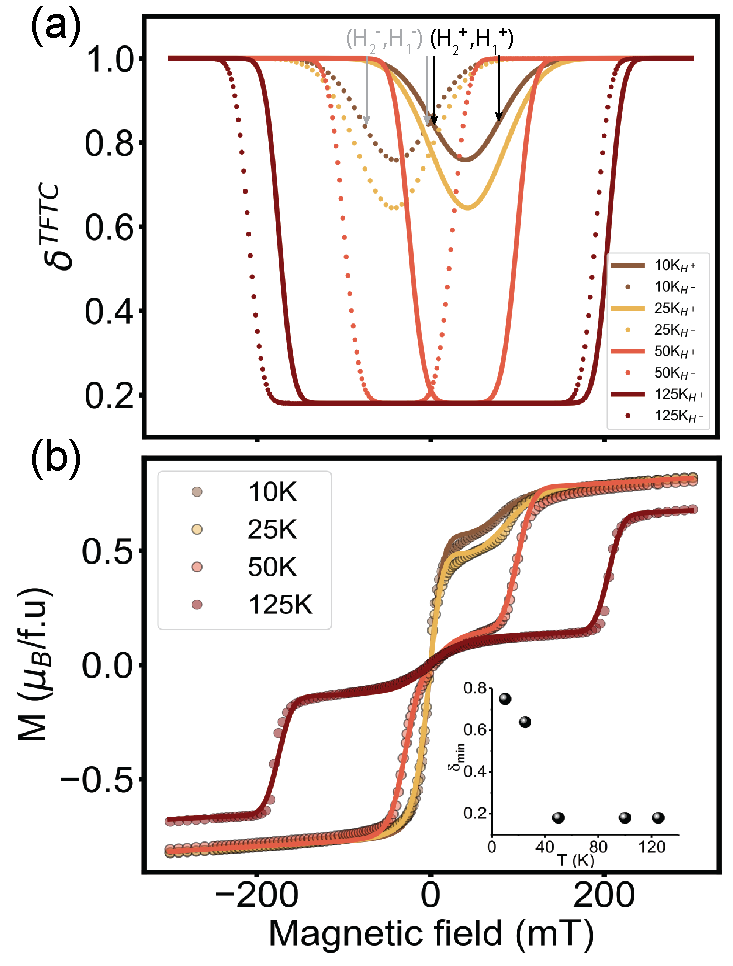}
    \caption{(a) Temperature- and field- dependent phase fraction of the TFTC structure, calculated using the $\delta(H,T)$ function in Eq. \ref{eq: delta} and parameters extracted from the fitting to the magnetisation data (Table \ref{tb:Delta}). Solid (dot) curves depict the fraction values change by increasing (decreasing) the field. (b) The models of magnetisation change by considering the $\delta(H,T)$ function at different temperature. Inset: Temperature dependence of the TFTC phase fraction $\delta_{min}(T)$ that `survives' during the switching process in the mixed phase at $H=0$.}
    \label{fig:6_Ratio_fit}
\end{figure}

To corroborate the above switching behaviour, we demonstrate here that a `minimal' model directly based on the magnetisation data is capable of fitting some of the fine details of our RXMS data in all polarisation channels. In order to achieve this, our model must include the relative fractions of the TFTC and LC phases. In the following we assume that no other phases are present, and that only the TFTC phase contributes to the in-plane magnetisation, which for a given magnetic domain we take to be independent of field (i.e. no change in the TFTC cone angle). 

We introduce the phase fraction parameter $\delta=V_{TFTC}/(V_{TFTC}+V_{LC})$ where $V_{TFTC}$ and $V_{LC}$ are the volumes of the two phases such that $\delta=1(0)$ for pure TFTC (LC) phases, respectively. The parameter $\delta$ is both field- and temperature dependent, and it is strongly hysteretic as a function of field -- in other words, the up-sweep (represented by $+$) and down-sweep (represented by $-$) functions $\delta^{+} (H,T)$ and $\delta^{-}(H,T)$ are different, being mirror image of each other through $H=0$ (see below).

The phase diagram in Figure \ref{fig:5_phase diagram} implies that the form of $\delta^{-}(H,T)$ vs $H$ is that of an inverted top hat, dropping from 1 at high fields to a temperature-dependent minimum value $\delta_{min}(T)$ upon decreasing the field, and rising again to 1 when the opposite field is increased. We have modelled the $\delta(H,T)$ function by a linear combination of back-to-back error functions:

\begin{align}
\label{eq: delta}
\delta(H,T) &= 1+\frac{(1-\delta_{min}(T))}{2}\left(\mathrm{erfc}\left[ \frac{\mu_0 H-\mu_0 H_2(T)}{w_c(T)}\right] \right. \nonumber \\
&\quad \left.-\mathrm{erfc}\left[ \frac{\mu_0 H-\mu_0 H_1(T)}{w_c(T)} \right] \right) 
\end{align}
where $w_c(T)$ represents the temperature dependent `speed of inter-conversion' of the phases (which may also depend on the field sweep rate), and $H_1(T)$ and $H_2(T)$ are the characteristic fields defining the sides of the top hat, which are also temperature dependent. We note that the symmetry of the hysteresis loop is maintained when $H_1^{-} \simeq -H_2^{+} \ll H_1^{+} \simeq -H_2^{-}$,  consistent with Figure \ref{fig:5_phase diagram}. The $\delta_{min}(T)$ and $w_c(T)$ functions can be estimated from the magnetometry measurements, while the characteristic fields can be read directly from the phase diagram in Figure \ref{fig:5_phase diagram}. 

The function $\delta(H,T)$ thus modelled is plotted in Figure \ref{fig:6_Ratio_fit} (a) for different temperatures and sweep directions, with the corresponding parameters listed in Table \ref{tb:Delta} -- only the values for field up-sweep (solid curves) are listed.

\begin{table}
  \centering
  \begin{tabular}{  |m{2.0cm}||m{1.0cm}|m{1.55cm}|m{1.55cm}|m{1.3cm}|}
      \hline 
      Temperature & $\delta_{min}$ &$\mu_0 H_1^+$(mT)&$\mu_0 H_2^+$(mT)& $w_c$(mT)\\
      \hline 
       \hfil 10 K &  \hfil 0.75 &  \hfil 70&  \hfil 10 &  \hfil 50 \\
      \hline
      \hfil 25 K & \hfil 0.64  & \hfil  80   & \hfil 5 &  \hfil 50\\
      \hline
      \hfil 50 K & \hfil 0.18 &  \hfil 100&  \hfil -25 &  \hfil 20\\
      \hline
      \hfil 125 K  & \hfil 0.18 &  \hfil 205&   \hfil -175 &  \hfil 18\\
      \hline
  \end{tabular}
  \caption{Parameters for the TFTC fraction function $\delta(H,T)$.}
  \label{tb:Delta}
\end{table}

Having modelled the field-dependent fractions of the TFTC and LC phases, we now consider the magnetisation of the TFTC phase alone, which switches upon rapid rotation about the $c$ axis. Following reference \citenum{Chmiel2019Magnetoelectric}, we describe the TFTC rotation angle about $c$ through a simple non-hysteretic function of field. Since the magnetometer measures the magnetization \emph{parallel} to the magnetic field, the  signal from the TFTC phase can be expressed as:  

\begin{equation}
\label{eq:par_mag}
      m_{\parallel}(H,T) = \lvert m(H,T)\rvert \sin \zeta
\end{equation}
where
\begin{equation}
\label{eq: par_magnetization}
  \zeta=\frac{\pi}{2} \tanh\left(\frac{\mu_0 H}{v_c(T)}\right)   
\end{equation}
is the cone rotation angle, whose low-field switching behaviour has been modelled with a hyperbolic cotangent function as we had done previously in ref. \citenum{Chmiel2019Magnetoelectric}.

In Eq. \ref{eq: par_magnetization}, $m(H,T)$ is the magnetisation of the TFTC phase, while the parameter $v_c(T)$ described the steepness of the magnetisation vs field curves (the `velocity of switching') and is itself temperature dependent. Note that, in fitting the magnetisation data, we have allowed $\lvert m(H,T)\rvert$ to be field dependent through the linear relation $m(H,T)= m(0,T) +\chi \ H$, which is necessary for an accurate fit of the high-field behaviour in fig. \ref{fig:6_Ratio_fit}.  At the microscopic level, this would correspond to a field-induced reduction of the TFTC cone angle. By combining the magnetisation rotation and phase fraction information, we obtain a single equation to fit the temperature- and field-dependent measured magnetisation data $m_\mathrm{meas.}(H,T)$:

\begin{equation}\label{eq:SQUID_model}
  m_\mathrm{meas.}(H,T) = \delta(H,T) \; m_{\parallel}(H,T)
\end{equation}
where $m_{\parallel}(H,T)$ is from Eq. \ref{eq:par_mag}. 

This model provides a satisfactory fit to the magentometry data, as shown in Figure \ref{fig:6_Ratio_fit} (b), with the fitting parameters of equation \ref{eq:SQUID_model} reported in Table \ref{table:squid}.   
\begin{table}
  \begin{tabular}{ |m{2.6cm}||m{1.3cm}|m{1.3cm}|m{1.3cm}|m{1.3cm} |}
      \hline 
      Temperature & \hfil 10 K &\hfil 25 K&\hfil 50 K& \hfil 125 K\\
      \hline 
       $v_c(T)$ (mT)   &  \hfil 25    & \hfil 25&  \hfil 50 & \hfil 55 \\
      \hline
       $\lvert m(0,T)\rvert$ ($\mu_B$/f.u)   &  \hfil 0.78    & \hfil 0.77 & \hfil 0.75  & \hfil 0.66 \\
      \hline
       $\chi$ (10$^{-4}\mu_B/mT$)   &  \hfil 3.53    & \hfil 3.63&  \hfil 3.87 & \hfil 4.60 \\
      \hline
   \end{tabular}
   \caption{Temperature dependent parameters for SQUID switching model in Eq. \ref{eq:par_mag}.}
   \label{table:squid}
\end{table}
The minimum TFTC phase fraction, $\delta_{min}$, is plotted as a function of temperature in the inset of panel \ref{fig:6_Ratio_fit} (b) (see also Table \ref{tb:Delta}).  $\delta_{min}$~ drops suddenly around 50 K, indicating that the TFTC becomes unstable (rather than metastable) in zero fields above this temperature. The susceptibility $\chi$ can be taken as temperature-independent, while $\lvert m(0,T)\rvert$ changes little in this temperature range, since the magnetic ordering temperature is above room temperature.

\begin{figure*}
  \centering
  \includegraphics[width= \textwidth]{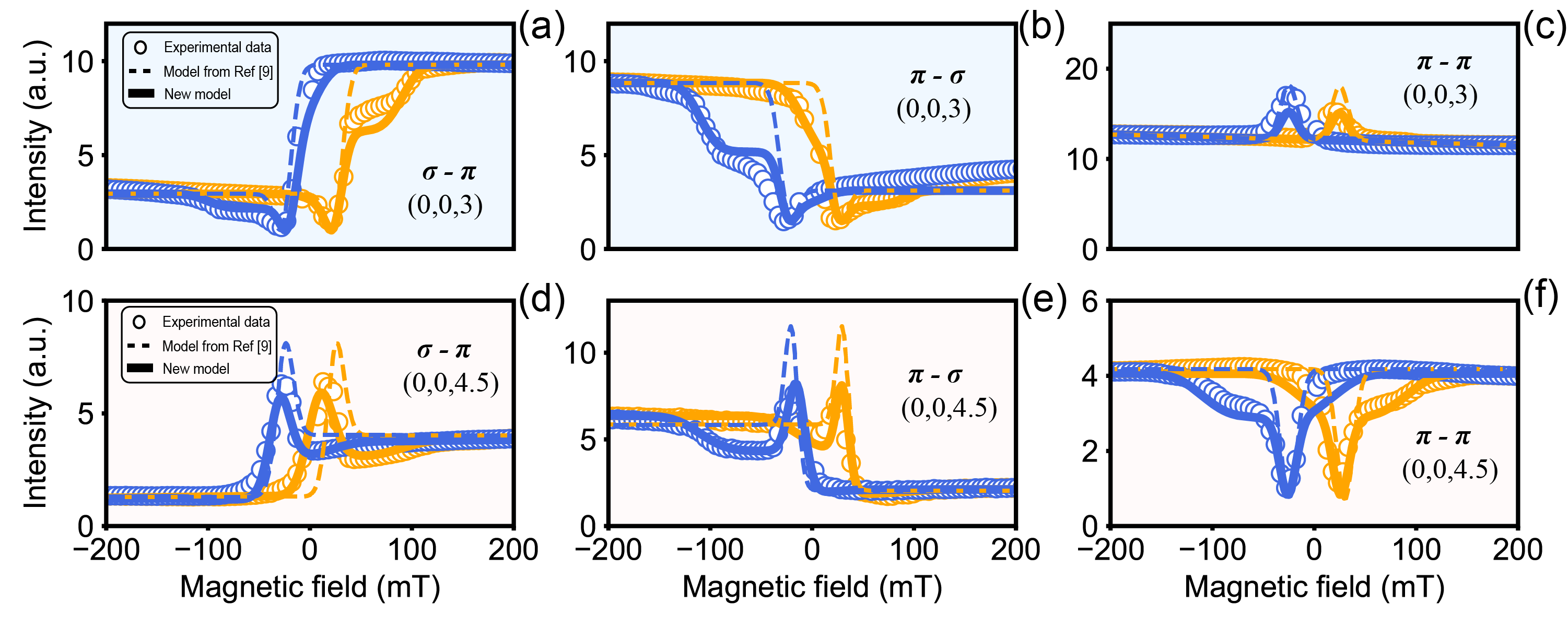}
  \caption{Calculated (dots) and measured (lines) scattering intensity for the (0,0,3) (panels (a)-(c))and (0,0,4.5) (panels (d)-(f)) reflections, in three linear polarization channels during the magnetic polarity switching process at 34 K. Yellow/blue lines and symbols are for field up-sweeps/down-sweeps, respectively.  Dashed lines are calculated using the model in reference \citenum{Chmiel2019Magnetoelectric}, while solid lines are calculated using the model described in the present paper, which takes into account the TFTC phase fraction.}
  \label{fig:7_Linear_polar_fit}
\end{figure*}

The parameters obtained from fitting the magnetometry data can now be used as a starting point to refine the modelling of our RXMS data through the magnetic field switching transition. Very similar data were previously presented in ref. \citenum{Chmiel2019Magnetoelectric}, which also contains a detailed analysis of the magnetic scattering. Very briefly, RXMS data were collected at the Fe $L_{3}$ edge (E=707 eV) on the $(0,0,3)$ and $(0,0,4.5)$ reflections, with linear incident X-ray polarisation, $\sigma$ or $\pi$, and with an analyser selecting the final linear polarisation to be detected, $\sigma'$ or $\pi'$. We measured field-dependent reflection intensities at T = 40 K in the three channels $\sigma-\pi'$, $\pi-\sigma'$, and $\pi-\pi'$, which are most sensitive to the components of the magnetic structure (see ref. \citenum{Chmiel2019Magnetoelectric} for details). The $(0,0,4.5)$ reflection is purely magnetic, while the $(0,0,3)$ reflection has a Thomson (structural) component in the $\pi-\pi'$ channel but not in the other two channels which are purely magnetic. In our geometry, the $\sigma-\pi'$, $\pi-\sigma'$ channels are sensitive to the angular momentum components in the scattering plane, i.e., $a^*$ (along the magnetic field) and $c$, whereas the $\pi-\pi'$ channel is sensitive to the $b$ component (orthogonal to the field), and its magnetic contribution is only activated during the field rotation.  In ref. \citenum{Chmiel2019Magnetoelectric}, the magnetic scattering was modelled using a modified version of the `standard' TFTC phase, which was found to be necessary to explain the RXMS. This m-TFTC model is also used as a basis to fit the data in the present paper.
In ref. \citenum{Chmiel2019Magnetoelectric}, RXMS data were analysed under the assumption of a rigid rotation of the TFTC phase and no change in phase fraction, which yielded a qualitatively acceptable agreement with the experimental observations. In particular, the rigid TFTC rotation model successfully describes the fast change in the RXMS diffraction intensity in the different $I_{channels}$ near $H=0$, as the magnetic cone structure rotates rapidly and non-hysterically through 180$^\circ$ about the $c$ axis.  

In spite of the qualitative similarities, there are obvious differences between the experimental and calculated RXMS profiles in ref. \citenum{Chmiel2019Magnetoelectric}. In particular, the data display clear anomalies at intermediate magnetic fields that are not reproduced by the model. The low-temperature, intermediate-field anomalies observed in both magnetisation and polarisation data appear at very similar magnetic fields to those found in the RXMS data, strongly suggesting that they have the same origin. In principle, the LC phase is weakly ferromagnetic and therefore contributes to the $\Gamma$-point magnetic signal, but in practice the magnetic moment from magnetometry is extremely small\cite{Hajime2009Two,ishiwata2010neutron,Koksis_2020} ($\approx 0.2 \mu_B$/f.u. from page 44 in \cite{chmiel2018a}) and no such scattering, with either linear or circular polarisations, was previously detected. Hence, to good approximation, the TFTC phase is the only one contributing to magnetic scattering at $(0,0,3)$, and uniquely contributes to the magnetic scattering at $(0,0,4.5)$. With this assumption, a minimal revised model is obtained simply by multiplying the calculated intensities in each channel from reference \citenum{Chmiel2019Magnetoelectric} by the TFTC phase fraction $\delta(H,T)$ as deduced above (N.B. we use $\delta(H,T)$ determined at 25 K). The intensities in each channel, for $\Gamma$ or $Z$ point diffraction peaks, are therefore:

\begin{widetext} \label{eq:appdxII_5}
  \begin{equation}
      \begin{split}
          I_{\Gamma/Z}^{\sigma-\pi}(H,T) &= \delta(H,T)\left|F^{(1)}[(M_{\Gamma/Z}^x \cos\zeta - M_{\Gamma/Z}^y \sin\zeta)\cos\theta + M_{\Gamma/Z}^z \sin\theta] \right|^2  \\
          I_{\Gamma/Z}^{\pi-\sigma}(H,T) &= \delta(H,T)\left|F^{(1)}[(M_{\Gamma/Z}^x \cos\zeta - M_{\Gamma/Z}^y \sin\zeta)\cos\theta - M_{\Gamma/Z}^z \sin\theta] \right|^2  \\
          I_{\Gamma/Z}^{\pi-\pi}(H,T) &= \delta(H,T)\left|F^{(0)} \cos 2\theta -i F^{(1)}[(M_{\Gamma/Z}^x \sin\zeta + M_{\Gamma/Z}^y \cos\zeta)\sin2\theta] \right|^2 \\
      \end{split}
  \end{equation}
\end{widetext}

where $\zeta$ is the rotation angle of the m-TFTC moment around the $c$ axis and is strongly field-dependent, $M_{\Gamma/Z}^x$, $M_{\Gamma/Z}^y$ and $M_{\Gamma/Z}^z$ are the components of the magnetic interaction vectors of the m-TFTC phase for the $(0,0,3)$ ($\Gamma$ point) and $(0,0,4.5)$ (commensurate $Z$ point) -- see \cite{Chmiel2019Magnetoelectric}, eq. B10, and $2 \theta$ is the scattering angle.  The internal parameters of the m-TFTC phase and the magnetic field dependence of $\zeta$ were kept as in \cite{Chmiel2019Magnetoelectric}, while the magnetic field dependence of $\delta$ was taken from the magnetometry data at 25 K, since no corresponding data at 40 K was available.

Figure \ref{fig:7_Linear_polar_fit} displays a comparison between the experimental RXMS data (filled circles), the original rigid-rotation model from reference \citenum{Chmiel2019Magnetoelectric} (short dash) and our new model including LC/TFTC intermediate mixed state (solid line). The top and bottom rows are the RXMS results obtained at the $(0,0,3)$ ($\Gamma$ point) and $(0,0,4.5)$ (commensurate $Z$ point) positions, respectively in the different cross-polarisation channels $\sigma-\pi'$, $\pi-\sigma'$ and $\pi-\pi'$. The new model is much closer to the experimental data, with the intermediate-field anomalies being well reproduced. 

As explained above, this `minimal' model has no adjustable parameters other than an overall scale factor, since all its parameters were taken either from our previous work \cite{Chmiel2019Magnetoelectric} or from our magnetometry data.  An obvious refinement of the model would entail introducing the field dependence of the main fan angle (referred to as $\phi$ in \cite{Chmiel2019Magnetoelectric}, Eq. B10), consistent with the non-zero magnetic susceptibility we observed in the magnetometry.  Although a refinement of these and other parameters would undoubtedly produce a better fit to the data, including the high-field slopes of panels (a) and (b) in \ref{fig:7_Linear_polar_fit}, we believe that our minimal model provides sufficient confirmation of the switching model without introducing additional complexity.

\section{Conclusions}
In conclusion, we have reported a detailed study of the field switching behaviour in the multiferroic Y-type hexaferrite \ce{Ba_{0.5}Sr_{1.5}Mg_{2}Fe_{12}O_{22}} (BSMFO), using SQUID magnetometry, magneto-current measurements and resonant X-ray magnetic scattering (RXMS) on single crystals that were poled using crossed magnetic and electric fields. Both magnetisation and electrical polarisation display an apparent multi-step hysteretic behaviour upon sweeping the external magnetic field at low temperatures.  Similar behaviour was previously reported, but not fully understood, in related materials. By combining these data sets, we are able to construct a model that closely reproduces all our observations.  In particular, we demonstrated that the multiferroic two-fan transverse cone (TFTC) phase, which is stable at low temperatures and at high magnetic fields, partially transforms into the longitudinal-cone (LC) phase upon field reversal, giving rise to an intermediate mixed state with reduced magnetisation and polarisation. At higher magnetic fields, the LC phase transforms back into TFTC yielding once again a pure state with maximum magnetisation and polarisation -- remarkably, without any need to re-apply an external bias.  We have also shown that this phenomenology is highly temperature-dependent, with the field of stability and phase fraction of the LC phase notably broadening/increasing upon warming.  Combining this new insight with our previous model of the TFTC cone rotation through $B=0$ produces a much better agreement with the RXMS data, yielding an essentially complete picture of the field-driven polarity switching in Y-type BSMFO, which can be likely extended to other multiferroic hexaferrites. 

\section*{Acknowledgments}

We acknowledge Beamline I10, Diamond Light Source (DLS) under Proposal No. SI17388 and No. SI20608, and thank P. Steadman and R. Fan for assisting the RXMS experiment. We thank J-C Lin and H. Jani for useful discussion. Jiahao Chen acknowledges the financial support from China Scholarship Council (CSC).

\bibliography{refs}

\end{document}